\documentclass[12pt]{article}
\usepackage[usenames,dvipsnames]{color}
\usepackage{graphicx}
\begin{document} 
\bibliographystyle{perso}
\date{}
\begin{center}
{\bf
INTEGRABLE LATTICE MAPS: $Q_5$, A RATIONAL VERSION OF  $Q_4$ } \\
\smallskip
Claude M. Viallet\footnote{viallet@lpthe.jussieu.fr}\\
\smallskip
{\em LPTHE, Centre National de la Recherche Scientifique, UPMC Univ. Paris 06}\\
{\em Bo\^{\i}te 126   /   4 Place Jussieu, F-75252 PARIS CEDEX 05}
\end{center}

\section{Contents} 

We give a rational form of a generic two-dimensional ``quad'' map,
containing the so-called
$Q_4$ case~\cite{Ad98,Ni02,AdBoSu03,AdSu04,Hi05,AdBoSu07}, but whose
coefficients are free.  Its integrability is proved using the
calculation of algebraic entropy.

We first explain the setting, i.e. what are two dimensional lattice
maps on a square lattice (quad maps), and describe two characteristics
of integrability of such systems, respectively {\em Lax pair and
  consistency}~\cite{Ni02,BoSu02}, with the important (generic)
example $Q_4$, and {\em vanishing of algebraic
  entropy}~\cite{BeVi99,HiVi98,TrGrRa01,Vi06}, which, as we will show,
provides a natural generalisation of $Q_4$, baptised $Q_5$.  We
explain the factorization process of the iterates at the origin of the
vanishing of the entropy, and present some directions for further
investigations.

\section{The setting}

Consider a field $x$ defined on a two-dimensional square lattice: at
each vertex of the lattice, the value of $x$ is related to the value
at neighbouring vertices. The simplest possible relation links the
values of $x$ at the four corners of each elementary square plaquette
by a multilinear relation
\begin{eqnarray}
  & Q = p_{1} \cdot x \; x_1 \; x_{2} \; x_{12} + p_{2} \cdot x \; x_{1}
\; x_{2} + p_{3} \cdot x \; x_{1} x_{12} + p_{4} \cdot x_{1} \; x_{2}
\; x_{12} + p_{5} \cdot x_{} \; x_{2} \; x_{12} & \nonumber
\\ & + p_{6} \cdot x
\; x_{2} + p_{7} \cdot x_{1} \; x_{2} + p_{8} \cdot x_{2} \; x_{12} +
p_{9} \cdot x \; x_{1} + p _{10} \cdot x \; x_{12} + p_{11} \cdot
x_{1} \; x_{12} & \nonumber
\\ & + p_{12} \cdot x_{2}+ p_{13} \cdot x + p_{14}
\cdot x_{1} + p_{15} \cdot x_{12} + p_{16} = 0 &
\label{cond}
\end{eqnarray}
so that any of the four corner values can be rationally expressed in
terms of the three others.

\vskip -1truecm
\begin{center}
\setlength{\unitlength}{1300sp}%
\begingroup\makeatletter\ifx\SetFigFont\undefined%
\gdef\SetFigFont#1#2#3#4#5{%
  \reset@font\fontsize{#1}{#2pt}%
  \fontfamily{#3}\fontseries{#4}\fontshape{#5}%
  \selectfont}%
\fi\endgroup%
\begin{picture}(6024,6024)(2989,-6973)
{\color[rgb]{0,0,0}\thinlines
\put(4201,-2161){\circle*{212}}
}%
{\color[rgb]{0,0,0}\put(7801,-2161){\circle*{212}}
}%
{\color[rgb]{0,0,0}\put(4201,-5761){\circle*{212}}
}%
{\color[rgb]{0,0,0}\put(7801,-5761){\circle*{212}}
}%
{\color[rgb]{0,0,0}\put(4201,-961){\line( 0,-1){6000}}
}%
{\color[rgb]{0,0,0}\put(7801,-961){\line( 0,-1){6000}}
}%
{\color[rgb]{0,0,0}\put(3001,-2161){\line( 1, 0){6000}}
}%
{\color[rgb]{0,0,0}\put(3001,-5761){\line( 1, 0){6000}}
}%
\put(8101,-6400){\makebox(0,0)[lb]{\smash{{\SetFigFont{20}{24.0}{\rmdefault}{\mddefault}{\updefault}{\color[rgb]{0,0,0}$x_{1}$}%
}}}}
\put(3000,-1936){\makebox(0,0)[lb]{\smash{{\SetFigFont{20}{24.0}{\rmdefault}{\mddefault}{\updefault}{\color[rgb]{0,0,0}$x_{2}$}%
}}}}
\put(3200,-6400){\makebox(0,0)[lb]{\smash{{\SetFigFont{20}{24.0}{\rmdefault}{\mddefault}{\updefault}{\color[rgb]{0,0,0}$x$}%
}}}}
\put(7951,-1936){\makebox(0,0)[lb]{\smash{{\SetFigFont{20}{24.0}{\rmdefault}{\mddefault}{\updefault}{\color[rgb]{0,0,0}$x_{12}$}%
}}}}
\end{picture}%
\end{center}

We will be interested in {\em  global} properties of the
evolutions defined by this {\em infinitesimal} relation, as
well as {\em local} constraints (like consistency around the
cube or factorization properties).

\section{Integrability: Lax pair and consistency around the cube (CAC)}
Consider the archetypal case of discrete mKdV:
\begin{eqnarray*}
p_1 \;(x \; x_1- x_2 \; x_{12})+ p_2 \;( x \; x_2 - x_1 \; x_{12}) = 0
\end{eqnarray*}

It is possible to embed the two-dimensional cell into a three-dimensional one:
\begin{center}
\setlength{\unitlength}{2000sp}%
\begingroup\makeatletter\ifx\SetFigFont\undefined%
\gdef\SetFigFont#1#2#3#4#5{%
  \reset@font\fontsize{#1}{#2pt}%
  \fontfamily{#3}\fontseries{#4}\fontshape{#5}%
  \selectfont}%
\fi\endgroup%
\begin{picture}(7353,5355)(2101,-7861)
{\color[rgb]{0,0,0}\thinlines
\put(2401,-7561){\circle*{150}}
}%
{\color[rgb]{0,0,0}\put(2401,-3961){\circle*{150}}
}%
{\color[rgb]{0,0,0}\put(5101,-6661){\circle*{150}}
}%
{\color[rgb]{0,0,0}\put(6001,-7561){\circle*{150}}
}%
{\color[rgb]{0,0,1}\put(5101,-3061){\circle*{212}}
}%
{\color[rgb]{0,0,1}\put(6001,-3961){\circle*{212}}
}%
{\color[rgb]{0,0,1}\put(8701,-6661){\circle*{212}}
}%
{\color[rgb]{1,0,0}\put(8701,-3061){\circle*{336}}
}%
{\color[rgb]{0,0,0}\put(2401,-7561){\line( 1, 0){3600}}
\put(6001,-7561){\line( 0, 1){3600}}
\put(6001,-3961){\line(-1, 0){3600}}
\put(2401,-3961){\line( 0,-1){3600}}
}%
{\color[rgb]{0,0,0}\put(2401,-3961){\line( 3, 1){2700}}
\put(5101,-3061){\line( 1, 0){3600}}
\put(8701,-3061){\line(-3,-1){2700}}
}%
{\color[rgb]{0,0,0}\put(2401,-7561){\line( 3, 1){2700}}
\put(5101,-6661){\line( 1, 0){3600}}
\put(8701,-6661){\line(-3,-1){2700}}
}%
{\color[rgb]{0,0,0}\put(5101,-6661){\line( 0, 1){3600}}
}%
{\color[rgb]{0,0,0}\put(8701,-6661){\line( 0, 1){3600}}
}%

\put(4000,-7861){\makebox(0,0)[lb]{\smash{{\SetFigFont{20}{24.0}{\rmdefault}{\mddefault}{\updefault}{\color[rgb]{1,0,0}$p_1$}%
}}}}
\put(3900,-4300){\makebox(0,0)[lb]{\smash{{\SetFigFont{20}{24.0}{\rmdefault}{\mddefault}{\updefault}{\color[rgb]{1,0,0}$-p_1$}%
}}}}
\put(1900,-5700){\makebox(0,0)[lb]{\smash{{\SetFigFont{20}{24.0}{\rmdefault}{\mddefault}{\updefault}{\color[rgb]{1,0,0}$p_2$}%
}}}}
\put(6000,-5700){\makebox(0,0)[lb]{\smash{{\SetFigFont{20}{24.0}{\rmdefault}{\mddefault}{\updefault}{\color[rgb]{1,0,0}$-p_2$}%
}}}}
\put(3500,-6900){\makebox(0,0)[lb]{\smash{{\SetFigFont{20}{24.0}{\rmdefault}{\mddefault}{\updefault}{\color[rgb]{1,0,0}$p_3$}%
}}}}
\put(3250,-3300){\makebox(0,0)[lb]{\smash{{\SetFigFont{20}{24.0}{\rmdefault}{\mddefault}{\updefault}{\color[rgb]{1,0,0}$-p_3$}%
}}}}
\put(6500,-2900){\makebox(0,0)[lb]{\smash{{\SetFigFont{20}{24.0}{\rmdefault}{\mddefault}{\updefault}{\color[rgb]{1,0,0}$p_1$}%
}}}}
\put(8800,-4600){\makebox(0,0)[lb]{\smash{{\SetFigFont{20}{24.0}{\rmdefault}{\mddefault}{\updefault}{\color[rgb]{1,0,0}$p_2$}%
}}}}
\put(7000,-7450){\makebox(0,0)[lb]{\smash{{\SetFigFont{20}{24.0}{\rmdefault}{\mddefault}{\updefault}{\color[rgb]{1,0,0}$-p_3$}%
}}}}
\put(7000,-4000){\makebox(0,0)[lb]{\smash{{\SetFigFont{20}{24.0}{\rmdefault}{\mddefault}{\updefault}{\color[rgb]{1,0,0}$p_3$}%
}}}}

\put(2000,-7861){\makebox(0,0)[lb]{\smash{{\SetFigFont{20}{24.0}{\rmdefault}{\mddefault}{\updefault}{\color[rgb]{0,0,0}$x$}%
}}}}
\put(6001,-7861){\makebox(0,0)[lb]{\smash{{\SetFigFont{20}{24.0}{\rmdefault}{\mddefault}{\updefault}{\color[rgb]{0,0,0}$x_1$}%
}}}}
\put(4801,-2761){\makebox(0,0)[lb]{\smash{{\SetFigFont{20}{24.0}{\rmdefault}{\mddefault}{\updefault}{\color[rgb]{0,0,0}$x_{23}$}%
}}}}
\put(8701,-2761){\makebox(0,0)[lb]{\smash{{\SetFigFont{20}{24.0}{\rmdefault}{\mddefault}{\updefault}{\color[rgb]{0,0,0}$x_{123}$}%
}}}}
\put(2101,-3661){\makebox(0,0)[lb]{\smash{{\SetFigFont{20}{24.0}{\rmdefault}{\mddefault}{\updefault}{\color[rgb]{0,0,0}$x_{2}$}%
}}}}
\put(4500,-6511){\makebox(0,0)[lb]{\smash{{\SetFigFont{20}{24.0}{\rmdefault}{\mddefault}{\updefault}{\color[rgb]{0,0,0}$x_{3}$}%
}}}}
\put(8851,-6511){\makebox(0,0)[lb]{\smash{{\SetFigFont{20}{24.0}{\rmdefault}{\mddefault}{\updefault}{\color[rgb]{0,0,0}$x_{13}$}%
}}}}
\put(5701,-3661){\makebox(0,0)[lb]{\smash{{\SetFigFont{20}{24.0}{\rmdefault}{\mddefault}{\updefault}{\color[rgb]{0,0,0}$x_{12}$}%
}}}}
\end{picture}%
\end{center}
where one imposes a similar relation to all faces (the same for
opposite faces).
\begin{eqnarray*}
p_i \; ( x \; x_i - x_j \; x_{ij} )+p_j \; ( x \; x_j - x_i \;
x_{ij})=0 ,\qquad  i,j = 1,2,3
\end{eqnarray*}

The higher dimensional system is compatible, i.e.  {\em the
value of $x_{123}$ is independent of the way it is calculated}.  This
is called consistency around the cube (CAC).

The major output of CAC is that it ensures the existence of a Lax
pair, which is accepted as a proof of integrability~\cite{Ni02,BoSu02}.

\section{Consistency around the cube: $Q_4$} 
While the defining plaquette  relation is written on one cell, and is
thus {\em infinitesimal}, the CAC relation is written on a loop of
cells, and is a {\em local} relation.

It is  a very constraining equation, and is not easy to manipulate: if one
takes the most general form of the defining relation $Q$, the
expressions of $x_{123}$ get quite difficult to handle, they are big.

We will be interested in the generic solution of CAC, i.e. the Adler
solution~\cite{Ad98}.  Its form has been improved by
Nijhoff\cite{Ni02}, and by Hietarinta~\cite{Hi05}. It was shown to be
the generic solution of CAC by
Adler-Bobenko-Suris~\cite{AdBoSu03,AdSu04,AdBoSu07}.  The solution was
called $Q_4$.  Its different avatars are respectively:

Adler's form:
\begin{eqnarray*} \hskip -2truecm
 & k_0 \; x \; x_1 \; x_2 \;x_{12} - k_1 ( x \; x_1 \; x_2 + x _1 \;
x_2 \; x_{12} + x \; x_2 \; x_{12} + x \; x_1 \; x_{12} ) + k_2 ( x \;
x_{12} + x_1 \; x_2 ) & \\ &- k_3 ( x \; x_1 + x_2 \; x_{12} ) - k_4 (
x \; x_2 + x_1 \; x_{12} ) + k_5 ( x + x_1 + x_2 + x_{12}) + k_6 =0 &
\\ & \hskip -1truecm \mbox{ with } k_0 = \alpha+\beta, \quad k_1 =
\alpha \nu + \beta \mu, \quad k_2 = \alpha \nu^2 + \beta \mu^2 , \quad
k_5 = {g_3\over{2}} k_0 + {g_2\over{4}} k_1 , \quad k_6 =
{g_2^2\over{16}} k_0 + {g_3} k_1, & \\ & k_3 =
{{\alpha\beta(\alpha+\beta)}\over{2(\nu - \mu)}} -\alpha \nu^2 + \beta
( 2\mu^2 - {g_2\over{4}}), \qquad k_4 = {{ \alpha \beta ( \alpha
+\beta)}\over{ 2(\mu- \nu)}} - \beta \mu^2 + \alpha ( 2\nu^2 -
{g_2\over{4}}). & \\ & \mbox{ and } \alpha^2 = r(\mu) , \qquad \beta^2
= r(\nu), \qquad r(z) =4 \; {z}^{3} - {\it g_2}\,z-{\it g_3}
\end{eqnarray*}

Nijhoff's form:
\begin{eqnarray*} 
 &  A \left( \left( x-b \right) \left( x_2-b \right) - d
 \right) \left( \left( x_1-b \right) \left( x_{12}-b \right) - d
 \right)  & \\ &  + B \left( \left( x-a \right) \left(
 x_1-a \right) - e \right) \left( \left( x_2-a \right) \left( x_{12}-a
 \right) - e \right) = f &
\end{eqnarray*}
where $(a,A)$, $(b,B)$, $(c,C) = (b,B) - (a,A)$ on the curve
${Z}^{2}= r(z)$,
\begin{eqnarray*}
\mbox { and  } \qquad d = (a-b)\;(c-b) \qquad e = (b-a)\; (c -a), \qquad f =
A\; B\;C \left( a-b \right)
\end{eqnarray*}

Hietarinta's form:
\begin{eqnarray*}
& sn(\alpha) \; sn(\beta) \; sn( \alpha + \beta) ( k^2 \; x \; x_1 \;
x_2 \;x_{12} + 1) + sn( \alpha + \beta) ( x \; x_{12} + x_1 \; x_2 )&
\\ & - sn( \alpha) ( x \; x_1 + x_2 \; x_{12} ) - sn(\beta) ( x \;
x_{2} + x_1 \; x_{12} ) = 0 &
\end{eqnarray*}

All three forms are parametrized through elliptic functions.  What we
will see is that there is another form, where the parameters are free
of any constraint. To see that, we will use the notion of algebraic
entropy.

\section{Algebraic entropy}
The space of initial data of the evolutions defined by
relation~(\ref{cond}) is infinite dimensional: indeed initial data
ought to be given on a line which allows the calculation of the values
at all points of the lattice. The simplest possible choice is to take
a regular diagonal staircase going diagonally (for more details
see~\cite{Vi06}). We then have a notion of iteration of the evolution
map, by calculating the values on diagonals moving away from the
initial staircase. This defines a sequence of degrees $d_n$ in terms
of the initial data, and leads to the entropy
\begin{eqnarray*}
\epsilon = \lim_{n\rightarrow\infty}{1\over{n}}\;  log(d_n) .
\end{eqnarray*}

The outcome of our numerous experiments, as well as what we know for
maps~\cite{FaVi93,HiVi98} leads to the claim that {\em integrability
  of the lattice map is equivalent to the vanishing of its entropy}.

\section{$Q_5$}
Apply this calculation to $Q_4$.  The most general form of
(\ref{cond}) having the same symmetries as $Q_4$ is:
\begin{eqnarray}
  & a_{1} \; x_{}  x_{1}  x_{2}  x_{12}+ a_{2}  \; \left(   x 
    x_{2}  x_{12}+  x_{1}  x_{2}  x_{12}+  x_{}  x_{1}  x_{12}+ 
    x_{}  x_{1}  x_{2} \right)  + a_{3} \; \left(   x  x_{1}+ 
    x_{2}  x_{12} \right) & \nonumber \\ &  +\;  a_{4} \;  
 \left(   x  x_{12}+  x_{1}  x_{2}
  \right) + a_{5} \; \left(   x_{1}  x_{12}+  x_{}  x_{2} \right)
 + a_{6} \; \left(   x_{}+  x_{1}+  x_{2}+  x_{12} \right)
  +a_{7} =0  \label{unconst}&
\end{eqnarray}

Since we use computer algebra to evaluate the sequence of degrees, it
is much more efficient to work with integer coefficients.  It is easy to find
integer coefficients verifying the conditions fulfilled by $\{ a_1,
\dots, a_7 \}$. For example, choosing $r(z) = 4\; z^3 - 32 \; z + 4$
and the points $ (a,A) = (0,2), \quad (c,C)=(3,4), \quad (b,B) = (a,A)
\oplus (c,C) = ( -26/9, -2/27) $, we get the sequence $ \{ d_n\} = \{
1, 3, 7, 13, 21, 31, 43, 57, 73, 91, 111, \dots\}$, that is to say the
{\em quadratic growth}
\begin{eqnarray*}
d_n = 1 + n \; ( n-1)
\end{eqnarray*}

 But we may also take the above form {\em without any constraint on
   the coefficients $\{ a_1, \dots, a_7 \}$}.  {\em With arbitrary
   values of the parameters, we get the same quadratic growth as with
   constrained values}:
\begin{eqnarray*} 
\{ d_n\} = \{ 1, 3, 7, 13, 21, 31, 43, 57, 73, 91, 111, \dots\}
\end{eqnarray*}
 fitted with the generating function
\begin{eqnarray*}
g(s) = \sum_{n=0}^{\infty}d_n \; s^n = {\frac
{1+{s}^{2}}{\left(1-s\right)^3 }}, \qquad {\mbox{ and }}\qquad d_n =
1 + n \;( n-1)
\end{eqnarray*}
as we have checked with a number of randomly chosen parameters.  This
indicates {\em integrability of the  unconstrained  form}, with 7 free
homogeneous parameters (intersection of hyperplanes in the space of
multilinear relations).  This is what we call $Q_5$.

Remark: the sequence of degrees verifies a finite recursion relation $
d_n = 3 \; d_{n-1} - 3 \; d_{n-2} + d_{n-3} $ This means that the
global behaviour of the sequence degrees is dictated by a local
condition.

\section{Factorization}

To analyse the origin of the entropy vanishing, one has to examine the factorization process, which explains the degree drop.
Consider the corner \par
\begin{center}
\includegraphics[height=5cm]{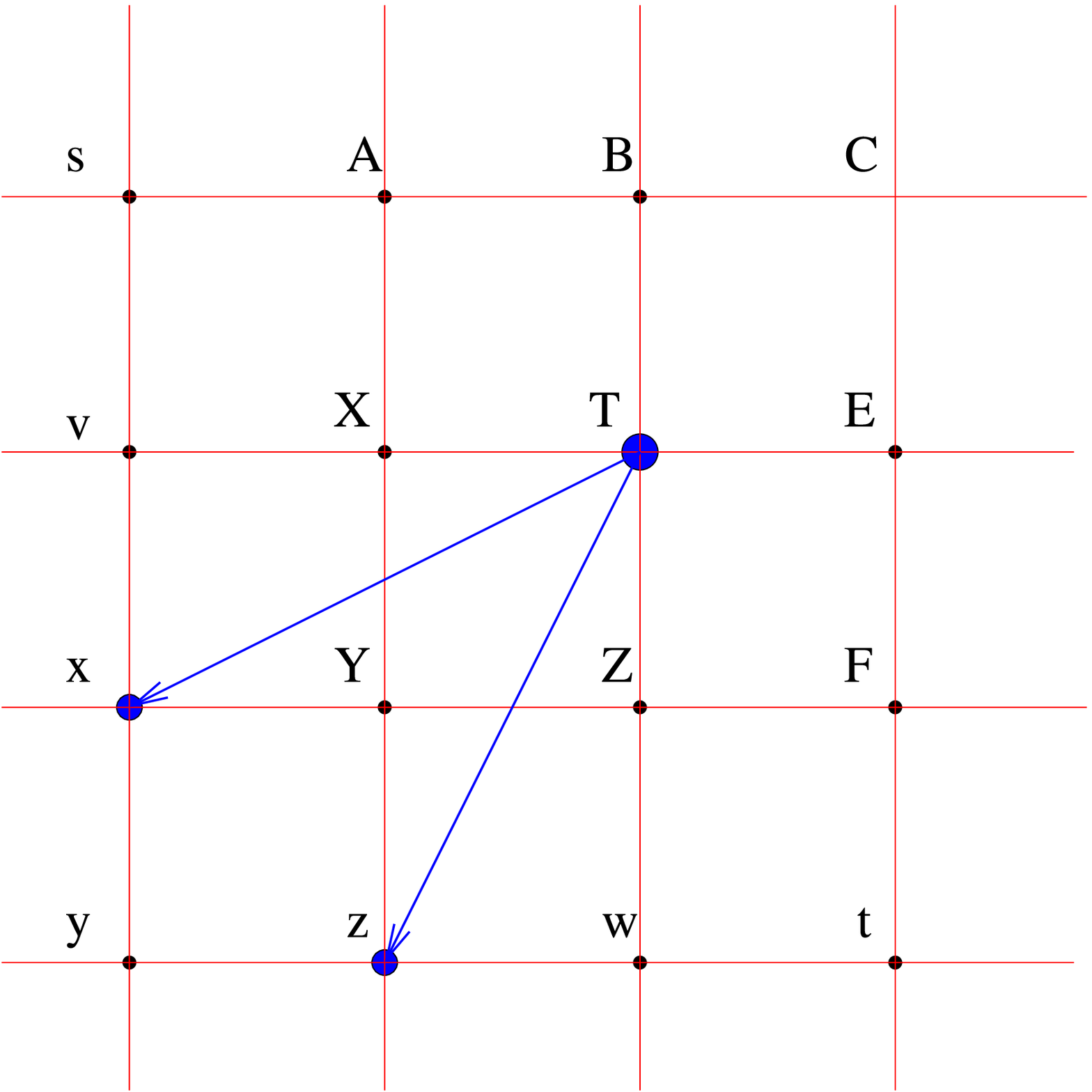} 
\end{center}
Suppose initial data( $s,v,x,y,z,w,t,\dots$) are given on the two axes.  
One can calculate the degree $d_{ij}$ at site $ij$and get, for $Q_5$
\begin{eqnarray*}
d_{ij} = 1 + 2 \; i \; j
\end{eqnarray*}
The diagonal degree growth is quadratic ( $ d_n = 1 + 2\; n^2 $) =
integrability

If we now evaluate $X$, $Y$, $Z$, $T$ for generic $Q$, with 16
independent coefficients, as in (\ref{cond}), we find:
\begin{eqnarray*}
& \mbox{deg}(Y) = 1+1+1 = 3, \qquad 
\mbox{deg}(X) =  \mbox{deg}(Z) = \mbox{deg}(Y)  +1+1 = 5  & \\ &
\mbox{deg}(T) = \mbox{deg}(X)+\mbox{deg}(Y)+\mbox{deg}(Z)  = 13 &
\end{eqnarray*}
What happens with  $Q_5$ is that there is a
factorization
\begin{eqnarray*}
T = { {H(x,z) \cdot P(x,y,z,u,v)}\over { H(x,z) \cdot Q(x,y,z,u,v)}} \simeq
{P\over{Q}}
\end{eqnarray*}
\begin{eqnarray*}
\mbox{deg}(T) = \mbox{deg}(X)+\mbox{deg}(Y)+\mbox{deg}(Z) {
- \mbox{deg}(H) } = 13 - 4 = 9
\end{eqnarray*}

The factor $H(x,z)$ is a bi-quadratic (elliptic) curve. It appears
naturally in the singularity analysis: suppose we look at the
elementary plaquette

\begin{center}
\setlength{\unitlength}{1300sp}%
\begingroup\makeatletter\ifx\SetFigFont\undefined%
\gdef\SetFigFont#1#2#3#4#5{%
  \reset@font\fontsize{#1}{#2pt}%
  \fontfamily{#3}\fontseries{#4}\fontshape{#5}%
  \selectfont}%
\fi\endgroup%
\begin{picture}(6024,6024)(2989,-6973)
{\color[rgb]{0,0,0}\thinlines
\put(4201,-2161){\circle*{212}}
}%
{\color[rgb]{0,0,0}\put(7801,-2161){\circle*{212}}
}%
{\color[rgb]{0,0,0}\put(4201,-5761){\circle*{212}}
}%
{\color[rgb]{0,0,0}\put(7801,-5761){\circle*{212}}
}%
{\color[rgb]{0,0,0}\put(4201,-961){\line( 0,-1){6000}}
}%
{\color[rgb]{0,0,0}\put(7801,-961){\line( 0,-1){6000}}
}%
{\color[rgb]{0,0,0}\put(3001,-2161){\line( 1, 0){6000}}
}%
{\color[rgb]{0,0,0}\put(3001,-5761){\line( 1, 0){6000}}
}%
\put(8101,-6336){\makebox(0,0)[lb]{\smash{{\SetFigFont{20}{24.0}{\rmdefault}{\mddefault}{\updefault}{\color[rgb]{0,0,0}$x$}%
}}}}
\put(3601,-1936){\makebox(0,0)[lb]{\smash{{\SetFigFont{20}{24.0}{\rmdefault}{\mddefault}{\updefault}{\color[rgb]{0,0,0}$z$}%
}}}}
\put(3576,-6336){\makebox(0,0)[lb]{\smash{{\SetFigFont{20}{24.0}{\rmdefault}{\mddefault}{\updefault}{\color[rgb]{0,0,0}$y$}%
}}}}
\put(7951,-1936){\makebox(0,0)[lb]{\smash{{\SetFigFont{20}{24.0}{\rmdefault}{\mddefault}{\updefault}{\color[rgb]{0,0,0}$Y$}%
}}}}
\end{picture}%
\end{center}
The relation $Q$ give a projective linear map $\varphi_{xz}: y
\longrightarrow Y$, whose inverse $\varphi^{-1}$ is projective
linear. The composed map $\varphi\cdot \varphi^{-1}$ is proportional to
the biquadratic $H(x,z)$, found in~\cite{AdBoSu07}.
\begin{eqnarray*}
\hskip -2truecm
 & H(x,z) = \left( p_{16}p_{10}-p_{15}p_{13}\right) + 
\left( -p_{8}p_{6}+p_{12}p_{5} \right) {x}^{2} + 
\left( p_{7}p_{3}-p_{2}p_{11} -p_{9}p_{4} +p_{14}p_{1} \right) {z}^{2}x & \\ & 
+ \left( -p_{6}p_{4}-p_{2}p_{8} + p_{7}p_{5}+p_{12}p_{1} \right) {x}^{2}z  
+ \left(-p_{4}p_{2}+p_{7}p_{1} \right) {x}^{2}{z}^{2} 
+ \left( -p_{11}p_{9} +p_{14}p_{3} \right) {z}^{2} & \\ & 
+ \left(-p_{2}p_{15}-p_{6}p_{11}+p_{7}p_{10} - p_{9}p_{8} +
p_{12}p_{3}+p_{16}p_{1}-p_{13}p_{4} + p_{14}p_{5} \right) xz & \\ & +
\left( p_{16}p_{3}-p_{13}p_{11} + p_{14}p_{10} - p_{9}p_{15} \right) z
+ \left( p_{16}p_{5}+p_{12}p_{10} - p_{13}p_{8} - p_{6}p_{15} \right) x &
\end{eqnarray*}

In the case of $Q_5$ the drop at $d_{22}$ is $13 -9 =4$. What
factorizes from the iterate is precisely equation of the bi-quadratic
$H(x,z)$. The elliptic curve of the known forms of $Q_4$ is lurking
there.

Remark: This does not account for the whole process, and higher degree
curves appear at later steps (total degree 16, degree 4 in $x$, $y$,
$z$, and bi-quadratic in $v$, $w$). What may however happen is that,
due to the specific form of the relation $Q$, it sufficient to ensure
that the first factorization happens to have them all.

This is spirit of a systematic analysis we have performed, for
quadratic relations, and with the additional hypothesis that factors
are made out of linear pieces (we know we will not find $Q_4$ this
way). This produced 80 a priori different models. We have run an
algebraic entropy test over those, and finally came out with a short
list of integrable cases, and a list of models with non-vanishing
entropy~\cite{HiVi07}.

Again some local structure, extending over a finite range of
elementary cells,  ensures a global property (integrability), as may be
seen form the existence of a finite recurrence relation on the
degrees.

\section{Conclusion and perspectives}

\begin{itemize}
\item
The three levels infinitesimal/local/global appear in the discrete
world.  In the setting we use, which is strongly constrained
(multilinearity of the elementary relation, birationality of the
evolution), a local property is good enough to ensure integrability.
\item About the rationality vs elliptic nature of the parametrization,
  the phenomenon is apparently the same as the one we
  saw~\cite{HiVi94} for the celebrated Baxter's solution of the
  Yang-Baxter equations. There exists a rational form of Baxter's
  R-matrix. It is gauge equivalent to the usual elliptic form, which
  reappears when one request a symmetric form of the solution.
\item
This phenomenon invites us to examine again the ``Yang-Baxter maps''
constructed from lattice maps~\cite{Dr92,Ve03,PaTo07} .
\item
Finally $Q_5$ will be useful if one wants to look at the possible
``de-autonomisations'' of $Q_4$.
\end{itemize}

\end{document}